\documentclass[a4p,11pt]{article}
\usepackage{amsmath,amssymb,bm,graphicx,subfigure}
\usepackage[margin=2cm]{geometry}
\usepackage{diagbox}
\usepackage[colorlinks=true]{hyperref}
\usepackage{algorithmic, algorithm}
\usepackage{textcomp}
\usepackage{xcolor}
\usepackage{mathtools}
\usepackage{microtype}
\usepackage{multirow}
\usepackage[symbol]{footmisc}
\newcommand{\tabref}[1]{Table~\ref{#1}} 
\newcommand{\figref}[1]{Fig.~\ref{#1}} 
\renewcommand{\eqref}[1]{Eq.~(\ref{#1})}
\newcommand{\secref}[1]{Section~\ref{#1}}

\hypersetup{colorlinks=true,linkcolor=darkblue,urlcolor=darkblue,anchorcolor=darkblue,citecolor=darkblue,breaklinks=true}
\definecolor{darkblue}{rgb}{0.0,0.0,0.8}
\begin{document}
%\begin{frontmatter}
\title{Damage identification using noisy frequency response functions based on topology optimization\thanks{This manuscript is the accepted version of the article: A. Saito, R. Sugai, Z. Wang, and H. Saomoto, ``Damage identification using noisy frequency response functions based on topology optimization,'' Journal of Sound and Vibration, 545, 117412 (2023). \url{https://doi.org/10.1016/j.jsv.2022.117412} \copyright\ 2023 The Author(s). Published by Elsevier Ltd. This manuscript version is made available
under the CC-BY-NC-ND 4.0 license
(\url{https://creativecommons.org/licenses/by-nc-nd/4.0/}).}}
%\author[meiji]{Akira~Saito\corref{cor1}}
%\ead{asaito@meiji.ac.jp}

\author{Akira~Saito\thanks{Corresponding author. 
Meiji University, Kawasaki, Kanagawa 214-8571, Japan. 
Email: asaito@meiji.ac.jp},
\and
Ryo~Sugai\thanks{Meiji University, Kawasaki, Kanagawa 214-8571, Japan.},
\and Zhongxu~Wang\thanks{Meiji University, Kawasaki, Kanagawa 214-8571, Japan.},
\and
Hidetaka~Saomoto\thanks{National Institute of Advanced Industrial Science and Technology, Tsukuba, Ibaraki 305-8567, Japan}
}
%
%\author[meiji]{Ryo~Sugai}
%\author[meiji]{Zhongxu~Wang}
%
%\author[aist]{Hidetaka~Saomoto}
%\address[meiji]{Meiji University, Kawasaki, Kanagawa 214-8571, Japan}
%\cortext[cor1]{Corresponding author. Tel:+81 44 934 7370.}
%\address[aist]{National Institute of Advanced Industrial Science and Technology, Tsukuba, Ibaraki 305-8567, Japan}
\date{}
\maketitle

\begin{abstract}
This paper proposes a robust damage identification method using noisy frequency response functions (FRFs) and topology optimization. We formulate the damage identification problem as an inverse problem of generating the damage topology of the structure from measured dynamic responses of the structure to given external dynamic loading. The method is based on the minimization of the objective function representing errors between measured FRFs of the structure obtained by experimental modal analysis, and those obtained by harmonic response analysis using finite element analysis. In the minimization process, material distribution, or the topology of the structure is varied and the optimal damage topology is identified as regions with no material assigned as a result of the minimization using the solid isotropic material with penalization (SIMP). In order to overcome the problems caused by the ill-posedness of the inverse problem, it is proposed that the least absolute shrinkage and selection operator (Lasso) regularization, or the penalization to the L1 norm of the design variable be applied to the original objective function. By applying Lasso regularization, the method is expected not only to eliminate spurious damaged regions but also to minimize the effect of measurement noises. This paper first presents the mathematical background and its numerical implementation of the proposed methodology. The method is then applied to the identification of a damage of cantilevered plates. The FRFs were experimentally obtained and the proposed method is applied. It is shown that the method successfully identifies the damage. 
\end{abstract}
\hrule
%\begin{keyword}
%\end{keyword}
%\end{frontmatter}
\section{Introduction}
Vibration-based damage detection methods for engineered structures have widely been used for many years, such as for civil structures~\cite{MeruaneHeylen2011,AvciEtAl2021} and mechanical structures~\cite{AdamsEtAl1978,GomesEtAl2019}. 
In general, the change in the vibration response of the structure for known external loadings is the consequence of the changes in the mechanical properties of the structure, such as stiffness and density. By exploiting this nature of the vibration response, if we treat the generation of damages on the structure as the change in its stiffness or density, we should be able to identify the damage by monitoring the changes in the vibration responses. 

{\color {black}{
There are two classes of promising vibration-based damage detection methodologies: the ones based on topology optimization algorithms~\cite{LeeEtAl2007,NiemannEtAl2010,SaomotoEtAl2015,NishizuEtAl2017,ZhangEtAl2017a}, such as the one proposed in this paper, and the ones based on data-driven machine learning based methods~(see, e.g.,~Refs.\cite{DackerMannEtAl2010, ChenEtAl2014, YuEtAl2018}).  
The latter approaches are preferred especially when the amount of available measurement data is sufficiently large. 
On the other hand, for the cases where the amount of available measurement data is not large, the ones based on optimization algorithms are advantageous because typically such methods use supposedly accurate mathematical models of the object of interest, 
which supplements the role of measurement data.  
%
% FE-based methods.
Regardless of the types of damage identification method, finite element method~(FEM) is widely utilized to represent the object and its damage to be identified~\cite{ChenEtAl2014,YeungSmith2005,GohEtAl2013,GonzalezEtAl2008}. From this, we can see that the use of FE models is an advantageous strategy. The proposed method uses topology optimization along with the FEM. 

%
%such vibration-based damage detection methods have been studied by many researchers. 
%which is capable of generating the topology of a design domain with which the performance of the domain is maximized. 
Damage detection using topology optimization~\cite{BendsoeKikuchi1988} with vibration response has been applied by some researchers. 
This class of methods can be applied even when a small number of measurement points are available. Moreover, in principle, its location, size, and shape of the damage can be identified simultaneously. }}
A pioneering contribution has been made by Lee {\it et al.}~\cite{LeeEtAl2007}, where they applied the topology optimization for damage detection using modal parameters. 
They employed 
%an objective function consisting of the differences in measured and computed resonant frequencies and those in measured and computed anti-resonant frequencies. 
{\color {black}{a penalty function as the objective function, which enforces the design variable to go toward either zero or one, along with multiple constraint equations about resonant and anti-resonant frequencies. With this formulation, it is possible to effectively handle many resonant and anti-resonant frequencies. }}
%
%Even though their results did not contain any study cases with real experimental data, since t
They succeeded in showing the effectiveness of their method to the detection of severely damaged structures, and the potential of the topology optimization in damage detection was demonstrated. 
It is noted that their results did not contain any study cases with real experimental data, and {\it spurious} damaged areas appeared in their results. 
Nieman {\it et al.}~\cite{NiemannEtAl2010} also employed topology optimization to damage identification problems based on the work of Lee {\it et al.}~\cite{LeeEtAl2007}.  They formulated the damage identification problem as the minimization of an objective function consisting of squared differences between measured and computed eigenvalues of the damaged structure and those between measured and computed frequency response functions (FRFs), along with a regularization of design variables. The proposed approach has been examined with purely numerical data and with experimentally-obtained modal parameters. Their method successfully identified internal and external damages generated on composite structures. However, they reported that spurious damaged regions appear with their method and sort of regularization is needed. 
%
%Even though it is not explicitly shown, it is stated that the regularization using the {\it $L_1$ norm} of the design variable is effective for their application. 
%A similar concept is employed in this paper. 
{\color{black}{It is stated that proper mass penalization using the design variable is necessary to remove spurious damaged areas in their application. As an example of such penalization, {\it $L_1$ norm} is employed in this paper.}}
Nishizu {\it et al.}~\cite{NishizuEtAl2017} also proposed a damage detection method based on the minimization of the weighted squared sum of the errors between the measured and computed eigenvalues of the structure. With their proposed method, a square notch in a cantilevered plate can be identified with satisfactory accuracy. However, their contribution only involves numerical experiments. 
Zhang {\it et al.}~\cite{ZhangEtAl2017a} proposed a level-set based approach for damage identification of continuous structures based on the dynamic responses. Their method successfully identifies damages with complicated topology by minimizing the weighted sum of errors between measured and computed natural frequencies. 

As seen in these articles, topology optimization using vibration-data, such as eigenfrequencies and FRFs has a potential as a damage-detection method. However, most methods have been validated over numerical data and their applicability to real structure is yet to be examined. Even with the method proposed by Nieman {\it et al.}~\cite{NiemannEtAl2010} that was examined with real experimental data, it suffered from spurious damaged regions. 

In this paper a damage-detection algorithm inspired by the previously proposed methods~\cite{LeeEtAl2007,NiemannEtAl2010} is proposed with an attempt to reduce the generation of spurious damages. 
{\color {black}{First, the effectiveness of the $L_1$ norm regularization method has been discussed with numerical experiments by Saomoto {\it et al}~\cite{SaomotoEtAl2015}, where the topology optimization was applied to the fault shape detection problem using static surface deformation of the ground. 
The authors then conducted preliminary studies to extend the method to the inverse problem of finding damages of cantilevered plates using frequency response functions~\cite{SugaiEtAl2020}. It was shown that the method works well with the frequency response functions. 
However, these were based on numerical experiments without any modeling and measurement errors. Moreover, experimental verification has never been conducted. }}
The experimental examination of the proposed method is given to prove its effectiveness in this paper. 

This paper is organized as follows. In \secref{sec:math}, the mathematical formulation of the proposed approach is described. 
%In \secref{sec:numerical_experiments}, numerical experiments with the proposed approach for damaged cantilevered plates using noisy FRFs where artificial measurement noises are injected, and the effects of measurement noises on the accuracy of the damage identification are investigated. 
In \secref{sec:experiments}, the proposed method is applied to measured FRFs and its validity is discussed. 
In \secref{sec:conclusions}, the conclusions of this paper are summarized. 

\section{Mathematical formulation}\label{sec:math}
In the proposed method, the damage identification problem is formulated as a design optimization problem where an objective function representing the discrepancies between the measured and computed quantities is defined, and is minimized with respect to a design variable, which is a variable that represents the material distribution. In the optimal design with the minimizer of the design variable, the damage is considered to be identified.
\subsection{Forward problem}\label{sec:forward_problem}
\begin{figure}[tb]
\centering
\includegraphics[height=5cm]{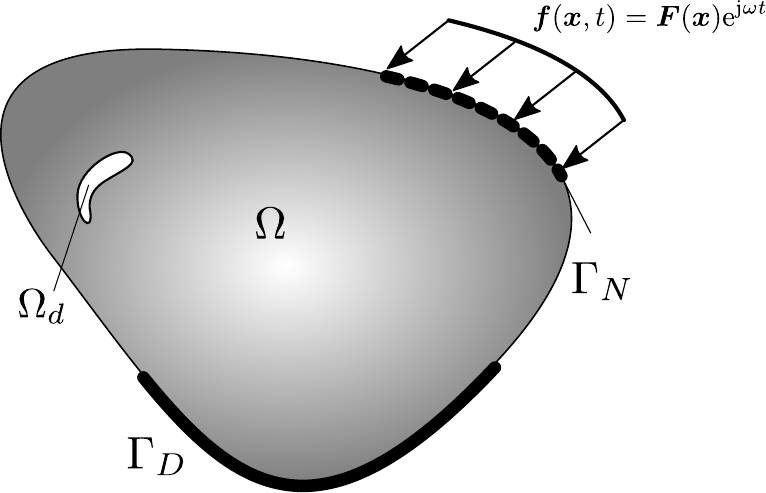}
\caption{Fixed design domain with a damaged region}\label{fig1}
\end{figure}
Let us assume that a three-dimensional domain of interest is occupied by an elastic media, and is denoted by $\bm{\Omega}\subset\mathbb{R}^3$, as shown in \figref{fig1}. 
The domain is assumed to contain damaged regions where the material is missing, which is denoted by $\bm{\Omega}_d$. 
Let us also assume that the domain is subject to harmonic loading with angular frequency $\omega\in\mathbb{R}$. This results in the dynamic deformation of the body. The displacement of any point on the domain is assumed to be small, such that any nonlinearity can be ignored. 
Also, damping is modeled as Rayleigh damping whose two components are proportional to mass and stiffness of the domain. 
Designating $\bm{x}\in\bm{\Omega}$ as the position vector of a material point in $\bm{\Omega}$ and $\bm{u}(\bm{x},t)\in\mathbb{R}^3$ as the displacement field. Then the governing equation for the forward problem of finding the displacement field for the given harmonic loading can be written in a strong form as follows: 
%https://www.comsol.com/blogs/part-1-modeling-the-harmonic-excitations-of-linear-systems/
%\begin{linenomath}
\begin{align}
\rho\ddot{\bm{u}}+\alpha\rho\dot{\bm{u}}&=\nabla\cdot\bm{\sigma}({\bm{u}})
+\beta\nabla\cdot\dot{\bm{\sigma}}(\bm{u})
\mbox{ in }\bm{\Omega},\label{forward_problem:eq1}\\
\bm{u}&=\bm{0}\mbox{ on }\Gamma_D,\\
\bm{\sigma}(\bm{u})\cdot{\bm{n}}&=\bm{f}(\bm{x},t)\mbox{ on }\Gamma_N,
\end{align}
where $\rho$ is the density, $\alpha$ and $\beta$ are the coefficients of the Rayleigh damping, $\Gamma_D$ denotes the Dirichlet boundary where displacement is prescribed, $\bm{f}(\bm{x},t)$ is the external harmonic force acting on the Neumann boundaries designated as $\Gamma_N$, and is defined as $\bm{f}(\bm{x},t)=\bm{F}(\bm{x}){\rm e}^{{\rm j}\omega t}$ where ${\rm j}=\sqrt{-1}$ and $\bm{F}(\bm{x})\in\mathbb{R}^3$ is the vector of forcing amplitudes. The stress tensor $\bm{\sigma}(\bm{u})$ is defined as a product between a fourth-rank elasticity tensor $\bm{C}$ and a strain tensor $\bm{\varepsilon}$: 
\begin{align}
\bm{\sigma}(\bm{u})&=\bm{C}:\bm{\varepsilon}(\bm{u}), \\
\bm{\varepsilon}(\bm{u})&=\frac{1}{2}\left\{\nabla\bm{u}+(\nabla\bm{u})^{\rm T}\right\}. 
 \end{align}
 where ":" denotes a double dot product. 
 Assuming that the displacement field is in steady-state, the displacement field can be written as 
 $\bm{u}(\bm{x},t)=\bm{U}(\bm{x},\omega){\rm e}^{{\rm j}\omega t}$, where $\bm{U}(\bm{x},\omega)\in\mathbb{C}^3$. Then, substituting $\bm{u}(\bm{x},t)$ into \eqref{forward_problem:eq1}, we obtain the following quasi-static governing equation for a given angular frequency $\omega$, 
\begin{align}
\left(-\omega^2+{\rm j}\alpha\omega\right)\rho{\bm{U}}&=\nabla\cdot\bm{\sigma}({\bm{U}})
+{\rm j}\beta\omega\nabla\cdot\bm{\sigma}({\bm{U}})
\mbox{ in }\bm{\Omega},\label{forward_problem:eq2}\\
\bm{U}&=\bm{0}\mbox{ on }\Gamma_D,\label{forward_problem:eq3}\\
\bm{\sigma}(\bm{U})\cdot{\bm{n}}&=\bm{F}(\bm{x})\mbox{ on }\Gamma_N.\label{forward_problem:eq4}
\end{align} 
%\end{linenomath}
The steady-state response can then be obtained by solving \eqref{forward_problem:eq2} subjected to Eqs.~(\ref{forward_problem:eq3}) and (\ref{forward_problem:eq4}). 
In this paper, this forward problem is solved by the FEM, and the frequency response functions (FRFs) are used as the solution of the forward problem. Since it is relatively easy to experimentally obtain the acceleration response due to harmonic loading by using widely-used accelerometers and force sensors, inertance or the accelerance is used as the FRFs. Namely, denoting the acceleration as $\bm{a}(\bm{x},t)$ and assuming that it is in steady-state, i.e., $\bm{a}(\bm{x},t)=\bm{A}(\bm{x},\omega){\rm e}^{{\rm j}\omega t}$, then $\bm{A}(\bm{x},\omega)=-\omega^2\bm{U}(\bm{x},\omega)$. 
It is also assumed that the force is applied to a single point whose location is denoted by $\bm{x}_e$, and the force is unidirectional along with a unit vector ${\bf e}_e$. Denoting the forcing amplitude as $F(\omega)\in\mathbb{C}$, then the vector of forcing amplitudes is written as $\bm{F}(\bm{x})=F(\omega)\delta(\bm{x}_e){\bf e}_e$ where $\delta$ denotes the Dirac delta function. The inertance is then defined as follows, 
\begin{equation}
\bm{H}(\bm{x},\omega)=\bm{A}(\bm{x},\omega)/F(\omega)=-\omega^2\bm{U}(\bm{x},\omega)/F(\omega). 
\end{equation}
Note that $F(\omega)$ can be set to 1 for the computation of numerical solutions, but in experiments, it is frequency-dependent. The details of the computation of FRFs with experimental data can be found in Ref.~\cite{Ewins2009} for instance. 
 %%%%%%%%%%%%%%%%%%%%%%%%%%%%%%
\subsection{Inverse problem}\label{sec:inverse_problem}
The design domain consists of a domain $\bm{\Omega}$ with subdomain $\bm{\Omega}_d$ that stands for the damaged domain where the material does {\it not} exist, and its complementary domain filled with material. 
Now, a characteristic function $\chi(\bm{x})$ that represents the material distribution is defined as follows: 
\begin{equation}
\chi(\bm{x})=
\left\{
\begin{array}{ccl}
0 &\mbox{for}&\forall\bm{x}\in\bm{\Omega}_d,\\
1 &\mbox{for}&\forall\bm{x}\in\bm{\Omega}\backslash\bm{\Omega}_d.
\end{array}
\right.
\label{eq1}
\end{equation}
Assuming that the solution of the forward problem \eqref{forward_problem:eq2} subjected to Eqs.~(\ref{forward_problem:eq3}) and (\ref{forward_problem:eq4}) is experimentally-obtained without knowing whether the domain contains a damage or not, the damage identification problem is defined as the problem of finding its material distribution for given solutions of the forward problem. 
The inverse problem is formulated as a minimization of a function that represents the discrepancy between the measured and computed solutions to the forward problem. 
%In this paper, frequency response function (FRF) is used as the solution of \eqref{forward_problem:eq2} subjected to Eqs.~(\ref{forward_problem:eq3}) and (\ref{forward_problem:eq4}). 
Namely, we define such a function to be minimized as the difference between the measured and computed FRFs as $J({\chi})$. 
The problem of the optimal material placement is then defined by the following minimization problem with a regularization: 
\begin{align}
\mathop{{\rm minimize}}_{\chi}\quad\!\!\!&Q(\chi):=J(\chi)+\lambda
L(\chi)
,\label{eq2}\\
\mbox{subject to}&: 0\leqslant\chi\leqslant1,\nonumber %\label{eq3}, 
\end{align}
where 
\begin{equation}
L(\chi)=\frac{\int_\Omega(1-\chi){\rm d}\Omega}{\int_\Omega{\rm d}\Omega}\label{lasso}. 
\end{equation}
$L(\chi)$ is a regularization term that penalizes the $L_1$ norm of $1-\chi$ with pre-defined regularization parameter $\lambda$. This term tends to suppress the increase of the value of $L_1$ norm of $1-\chi$, which means that it tends to fulfill the domain with material with $\chi=1$ as much as possible. 
This term is referred to as the least absolute shrinkage and selection operator, or Lasso~\cite{Tibshirani1996,Tibshirani2011}. 
{\color{black}{What we are proposing in this formulation is twofold: the suppression of spurious damaged areas, and the placement of material as much as possible with the assumption that the damaged regions to be identified are expected to be small, which stems from the engineering judgement that the damages of engineered structures should be identified before they grow. To achieve these objectives, the $L_1$ norm of $1-\chi$ is penalized. 
Also, as it was proposed in Ref.~\cite{LeeEtAl2007}, if we were to prefer solutions with clear boundary, double-well potential type objective functions, i.e., the norm of $\chi(1-\chi)$, would work because it enhance the binarization of the design variable. However, since the real damaged regions can consist of material distributions with intermediate values, such strategy to binarize the design variables was not employed in the proposed method. This leaves the capability to evaluate the degree of the damage by exploiting the intermediate values of the design variables.
}}

In the proposed method, Solid Isotropic Material with Penalization (SIMP)~\cite{BendsoeSigmund1999} is applied to represent the material distribution. Namely, Young's modulus $E$ and density $\rho$ of the body are interpolated as 
\begin{align}
\tilde{E}&=\chi^p E, \\
\tilde{\rho}&=\chi^q\rho, 
\end{align}
where the exponents $p$ and $q$ are pre-set penalty parameters, whose values are set to 3 and 1, respectively, as suggested in Ref.~\cite{BendsoeSigmund1999}. 
{\color{black}{When topology optimization is applied to the optimization of eigenfrequencies of engineered structures, it is known that spurious localized vibration modes with low frequencies tends to appear~\cite{Tcherniak2002,ZarghamEtAl2016}. To eliminate such spurious vibration modes, higher mass penalization can be used because it results in increasing the eigenfrequencies of the localized vibration modes. By using larger penalty factor for the mass density than that for the stiffness, it suppresses inertial effect stronger than the stiffness effect with very low material distribution, or almost void material, resulting in increasing the eigenfrequencies for the localized modes. However, for the inverse problems of our interest, since the domain tends to be filled with material as much as possible because of the Lasso regularization, the issue of localized vibration mode is limited, because the regions with almost void material disappear relatively quickly during the iteration. In addition, since frequency response functions are used in the proposed method, as long as low frequency bands that may contain spurious vibration modes are not used in the inverse problems, generation of spurious localized eigenmodes is not an issue for the proposed method. Therefore, relatively low penalty factor for the mass density has been employed in this work.
}}
\subsection{The objective function for the damage identification problem}\label{sec:objective_function}
As the objective function to be minimized, two objective functions are examined in this study. Assuming that the FRFs are obtained at $N$ measurement points for $n_f$ frequencies, vectors of computed and measured FRFs are defined as, ${\bf H}(\chi,\omega_j)=[\bm{H}(\chi,\bm{x}_1,\omega_j),\dots,\bm{H}(\chi,\bm{x}_N,\omega_j)]^{\rm T}$ and $\tilde{\bf H}(\omega_j)=[\tilde{\bm{H}}(\bm{x}_1,\omega_j),\dots,\tilde{\bm{H}}(\bm{x}_N,\omega_j)]^{\rm T}$, respectively. 
With these quantities, an objective function based on the mean square error (MSE) is introduced: 
\begin{equation}
J_1(\chi)
%\frac{1}{n_f}\sum^{n_f}_{j=1}
%\left\lVert
%{\bf H}(\chi,\omega_j)-\tilde{{\bf H}}(\omega_j)
%\right\rVert^2_2
=\frac{1}{n_f}\sum^{n_f}_{j=1}
\left\lVert
{\bf E}(\chi,\omega_j)
\right\rVert^2_2
, \label{eq4}
\end{equation}
where $\lVert\cdot\rVert_2$ represents the 2-norm of a vector, and ${\bf E}(\chi,\omega_j)$ denotes the error between the measured and computed FRFs, i.e., 
\begin{equation}
{\bf E}(\chi,\omega)={\bf H}(\chi,\omega_j)-\tilde{{\bf H}}(\omega_j). 
\end{equation}
$J_1(\chi)$ evaluates the point-wise error between the measured and computed FRFs. 
Requiring this to be minimized at an every single point on the body is a {\it strong} requirement, which may not be possible especially when noisy measurement data are involved. 
Therefore, a somewhat relaxed objective function is introduced, which is based on the mode assurance criterion (MAC)~\cite{Allemang2003} and is defined as: 
\begin{equation}
J_2(\chi)=\sum^{n_f}_{j=1}
M(\chi,\omega_j), 
\end{equation}
where 
\begin{equation}
M(\chi,\omega_j)=1-
\frac{
\left\{{\bf H}(\chi,\omega_j)\cdot \tilde{{\bf H}}(\omega_j)\right\}^2
}
{
\lVert{\bf H}(\chi,\omega_j)\rVert^2_2
\lVert\tilde{{\bf H}}(\omega_j)\rVert_2^2
}.\label{eq5}
\end{equation}
The second term on the right-hand side of \eqref{eq5} represents the MAC between ${\bf H}(\chi,\omega_j)$ and $\tilde{\bf H}(\omega_j)$, which takes the value from zero to one. 
If the value of MAC is equal to zero, the two vectors are orthogonal to each other, whereas if it is equal to one, then they are parallel to each other. 
In other words, if ${\bf H}(\chi,\omega_j)$ and $\tilde{\bf H}(\omega_j)$ are similar to each other, then $M(\chi,\omega_j)$ becomes close to 0, whereas if they are not similar to each other, then $M(\chi,\omega_j)$ becomes close to 1. 
Therefore, minimizing the sum of $M(\chi,\omega_j)$ is expected to produce the material distribution with which ${\bf H}(\chi,\omega_j)$ becomes similar to $\tilde{\bf H}(\chi,\omega_j)$, which is the FRF of the structure with the damage to be identified. 
%On the other hand, 
$J_2(\chi)$ evaluates the similarity of ${\bf H}(\chi,\omega_j)$ and $\tilde{\bf H}(\chi)$ in the sense of inner product, which indicates a {\it weak} requirement in comparison with \eqref{eq4}. 
In the following, both functions are examined, i.e., minimization problem defined in \eqref{eq2} is solved with $J(\chi)=J_1(\chi)$ and $J(\chi)=J_2(\chi)$. 

It is noted that many existing vibration-based damage identification methods, such as Refs~\cite{LeeEtAl2007,ZhangEtAl2017a}  utilize the minimization of the errors in the computed and the measured natural frequencies. The use of FRFs is advantageous over the use of natural frequencies because the process of curve-fit to obtain natural frequencies can be avoided, which is error-prone. 

\subsection{Analysis procedure}\label{sec:analysis_procedure}
Using the objective functions defined in Section \ref{sec:objective_function}, the solutions to the inverse problem are sought. The analysis procedure is shown in Algorithm~\ref{alg1}. First, the initial material distribution is set. In the analyses shown in this paper, $\chi_0(\bm{x})=1$. The value of the regularization parameter is also set to a given value. 
The forward problem Eq.~(\ref{forward_problem:eq2}) subject to Eqs.~(\ref{forward_problem:eq3}) and (\ref{forward_problem:eq4}) is then solved by an FEA. The objective function $L(\chi)$ is evaluated by computing $J(\chi)$ and the regularization term. Note that $J(\chi)$ is either $J_1(\chi)$ or $J_2(\chi)$ in this paper. 
The values of the objective and its gradient are evaluated, and the termination criteria are satisfied, then the converged solution is obtained. Otherwise, the material distribution $\chi(\bm{x})$ is updated and the process is repeated until the termination criteria are satisfied, or it reaches the pre-set maximum number of iterations. 
\begin{algorithm}[bt]
\begin{algorithmic}[1]
\caption{Procedure of the inverse analysis}\label{alg1}
\STATE{Set the initial material distribution $\chi(\bm{x})=\chi_0(\bm{x})$}
\STATE{Set the value of the regularization parameter $\lambda$}
\WHILE{$i\leqslant$Maximum number of iteration}
\FOR{$j=1,\dots,n_f$}
\STATE{Solve the forward problem Eq.~(\ref{forward_problem:eq2}) subject to Eqs.~(\ref{forward_problem:eq3}) and (\ref{forward_problem:eq4}) for {\bf H}($\chi,\omega_j$)}
\ENDFOR
\STATE{$Q(\chi)\leftarrow{J}(\chi)+\lambda{L(\chi)}$}%\int_\Omega(1-\chi){\rm d}\Omega/\int_\Omega{\rm d}\Omega$}
\IF{Termination criteria based on $Q(\chi)$ and its gradient are satisfied}
\STATE{break}
\ELSE
\STATE{Update $\chi(\bm{x})$}
\ENDIF
\ENDWHILE
\end{algorithmic}
\end{algorithm}
%
%%%%%%%%%%%%%%%%%%%%%%%%%%%%%%%%%%%%%%%%%%%%%%%%%%%%%%%%%%%%%%%%%%%%%%%%
\section{Damage identification based on experimentally obtained frequency response functions}\label{sec:experiments}
This section provides the results of the application of the proposed method to a cantilevered plate with a notch. 
\subsection{Damaged plates with a notch}
\begin{figure}[bt]
        \centering
        \subfigure[Specimen A]{
                \includegraphics[scale=0.8]{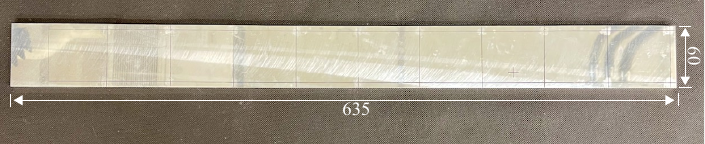}}
        \subfigure[Specimen B]{
        \includegraphics[scale=0.8]{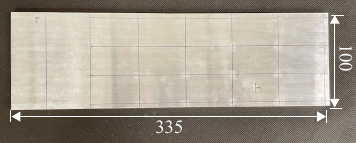}}
    \caption{Test specimen without damage}
    \label{exp_shape_nonDmg}
\end{figure}
In this study, a damage is defined as a loss of material. To make such artificially damaged plates, a notch is made on a rectangular plate. To examine the effects of aspect ratio on the identification results, two types of specimens with different aspect ratios were used. 
The photographs of the test specimen without the notch are shown in \figref{exp_shape_nonDmg}. 
The specimen A shown in Fig.~\ref{exp_shape_nonDmg}(a) has relatively high aspect ratio, whose dimension is 635mm$\times$60mm$\times$5mm. 
%On the other hand, t
The specimen B shown in Fig.~\ref{exp_shape_nonDmg}(b) has relatively low aspect ratio, whose dimension is 335mm$\times$100mm$\times$5mm. 
They are made of an aluminum alloy (A2017).% with density $\rho=2700\mathrm{kg/m}^3$, Young's modulus $70\mathrm{GPa}$, and Poisson's ratio $\nu=0.33$. 
\begin{figure}[bt]
        \centering
        \subfigure[Case A1: Specimen A with a notch on the short edge]{
        \includegraphics[scale=.9]{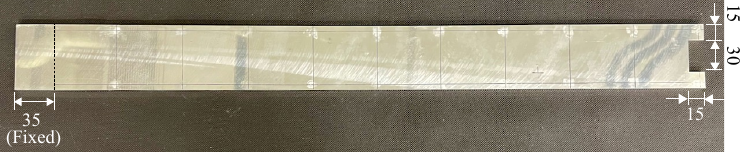}}        
        \subfigure[Case A2: Specimen A with a notch on the long edge]{
        \includegraphics[scale=.9]{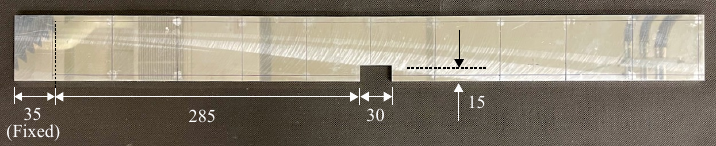}}
        \caption{Dimension and photograph of the specimen A with a notch}\label{exp_shape_A_Dmg}
\end{figure}
Figure \ref{exp_shape_A_Dmg} shows the dimension and photograph of the specimen A with a notch. The case where the notch is present on the short edge is designated as Case A1, whereas the notch is present on the long edge is designated as Case A2. 
\begin{figure}[tb]
\centering
\subfigure[Case B1: Specimen B with a notch on the short edge]{
\includegraphics[scale=0.8]{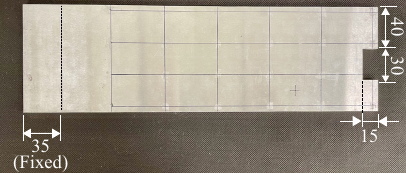}}
\subfigure[Case B2: Specimen B with a notch on the long edge]{
\includegraphics[scale=0.8]{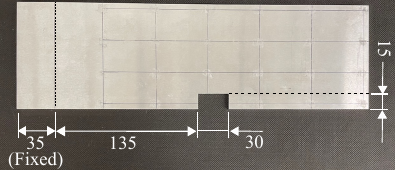}}
\caption{Dimension and photograph of the specimen B with a notch}\label{exp_shape_B_Dmg}
\end{figure}
Similarly, Fig.~\ref{exp_shape_B_Dmg} shows the dimension and photograph of the specimen B with a notch. This time, the case where the notch is present on the short edge is designated as Case B1, whereas the notch is present on the long edge is designated as Case B2. 
\subsection{Test procedure}
\begin{figure}[tb]
\centering
\subfigure[Photograph of the experimental setup]{\includegraphics[scale=1]{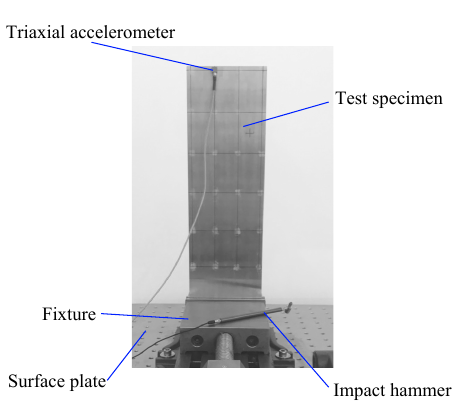}}
\subfigure[Measurement and excitation points]{\includegraphics[scale=.9]{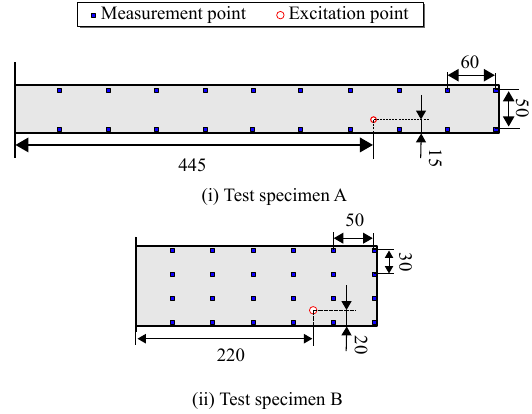}}
\caption{Experimental setup}
\label{exp_setup}
\end{figure}

This subsection provides the setup and the procedure of the experiments. In this study, FRFs are used for the damage identification. In particular, the inertance, or the accelerance is used, i.e., the ratio of the Fourier spectrum of the response acceleration and that of the input force needs to be experimentally obtained for the frequency range of interest. In this study, the impact hammer (GK-2110, Onosokki, Japan) is used for the excitation of the plate, and accelerometers (356A01, PCB Piezotronics, USA) were used for the measurement of the out-of-plane accelerations. 
Photograph of the test setup is shown in \figref{exp_setup}(a). As shown in the figure, the test specimen was clamped by a fixture
 %(E-9107 F100, Nabeya Bi-tech, Japan) 
 that is fixed on a surface plate to ensure that the fixed boundary condition is realized. 
The accelerometers were then mounted on the plate surface using a petro wax. 
% (080A24, PCB Piezotronics, USA). 
%
The output signals from both the accelerometers and the impact hammer were transferred to an FFT analyzer system (DT9837A, Measurement Computing, USA), and the FRFs were obtained. 

Next, test conditions are described. The measurement and the excitation points for the test specimens are shown in \figref{exp_setup}(b). 
The accelerometers were placed at equally-spaced points on the surface of the plates and measurements were conducted. The number of measurement points for the test specimen A and B were 20 and 24, respectively. 
For the FFT analyses, exponential window was used for the acceleration to reduce leakage errors, whereas the rectangular window was used for the force to reduce measurement noise. 
The experiments were conducted eight times and the averaged spectra were obtained. 
The FRFs were obtained for the test specimen A with the sampling frequency of 1~kHz and the measurement duration of 4~sec, whereas for the test specimen B with the sampling frequency of 2~kHz and the measurement duration of 2~sec. 
\subsection{Test results}\label{subsec:testresults}
\begin{figure}[bt]
        \centering
        \subfigure[Test specimen A]{\includegraphics[scale=.9]{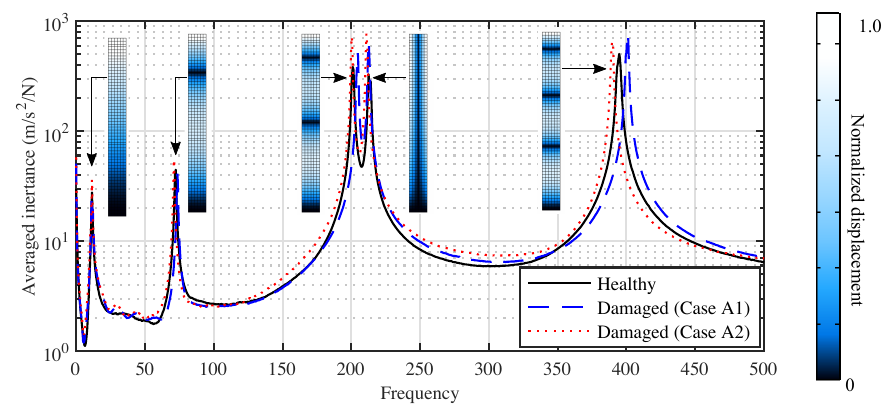}}
        \subfigure[Test specimen B]{\includegraphics[scale=.9]{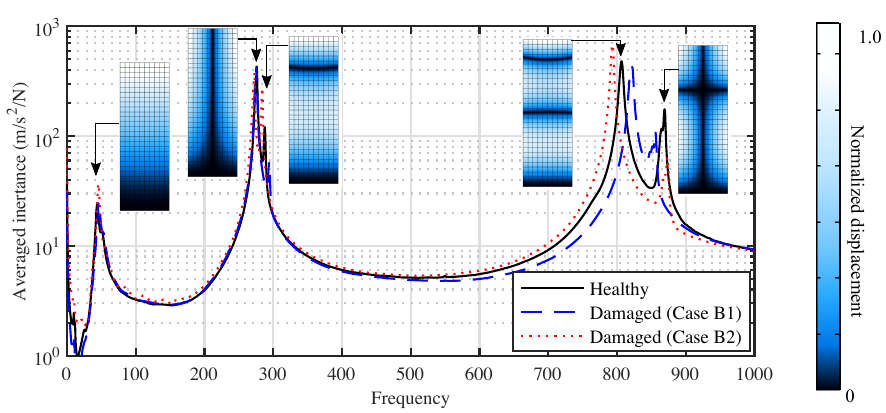}}
        \caption{Measured FRFs for the test specimens with numerically obtained mode shapes. {\color{black}{The FRFs are available at Mendeley Data (\href{http://dx.doi.org/10.17632/r3m5gmjsnv.1}{http://dx.doi.org/10.17632/r3m5gmjsnv.1}).}}}\label{fig6}
\end{figure}

The measured FRFs for the test specimen A and B are shown in \figref{fig6}(a) and \figref{fig6}(b), respectively. 
{\color{black}{The FRFs are available at Mendeley Data (\href{http://dx.doi.org/10.17632/r3m5gmjsnv.1}{http://dx.doi.org/10.17632/r3m5gmjsnv.1})}.}
Note that the FRFs are averaged for all measurement points.
 %and directions. 
 Also, the mode shapes of the healthy plates corresponding to the resonant peaks are shown in the graphs. The mode shapes were computed by COMSOL Multiphysics\textsuperscript{\textregistered}. 
As can be seen from the resonant peaks in \figref{fig6}(a), there are five modes below 500Hz for the test specimen A. The frequency deviations due to the damages are almost negligible for the first two modes. However, for the modes from three to five, as the frequency increases, the deviations in the frequency due to the damages grow, as can be seen in the FRFs. 
For the fifth mode at around 400Hz, we can see that the natural frequency decreases for the Case A2, whereas it increases for the Case A1, in comparison with that of the healthy case. 
The decrease in the natural frequency for the Case A1 agrees with the fact that the damage causes the decrease in the natural frequencies, especially for cracks~\cite{Dimarogonas1996}. This is because the influence of the damage on the decrease in the modal stiffness is larger than that in the modal mass for the Case A2, which results in the decrease in the natural frequency. 
On the other hand, the natural frequency increased for the Case A1 in comparison with that of the healthy case. This is because the decrease in the modal mass is larger than the decrease in the modal stiffness for the Case A1 for this mode, which results in the increase in the natural frequency. 
Furthermore, maximum values of the FRFs at the peaks also increase due to the damages for all damaged cases considered. This makes sense considering that the damages decrease the stiffness of the plates, resulting in the increases in the response. 

For the test specimen B, there are five modes below 1000Hz. Similar trends can be observed for this test specimen, i.e., as the natural frequencies increase, deviations in the frequencies due to the damages increase. 
Furthermore, the notch causes some of the natural frequencies to decrease or increase, depending on its influence on the modal mass and stiffness. 

%\clearpage
\subsection{Procedure of the inverse analysis}
% Explanation of the objective function
With the FRFs shown in Section \ref{subsec:testresults}, inverse analyses were conducted.  
The procedure is described as follows. 
First, the measured FRFs were used as $\tilde{\bf H}(\omega_j)$, for $j=1,\dots,n_f$ in \eqref{eq4} and (\ref{eq5}). 
Second, ${\bf H}(\chi,\omega_j)$ in \eqref{eq4} and (\ref{eq5}) were computed by solving the forward problem defined in \eqref{forward_problem:eq2} subjected to Eqs.~(\ref{forward_problem:eq3}) and (\ref{forward_problem:eq4}) using the FEM for a given material distribution $\chi$. 
Then the objective function in \eqref{eq2} was evaluated with a given $\lambda$. 
The initial value of $\chi$ was set to 1 everywhere. 
If it reached the minimum, or a certain termination criterion was satisfied, the obtained material distribution was considered as the damaged topology. Otherwise, the material distribution $\chi$ was updated and the forward problem \eqref{forward_problem:eq2} subjected to Eqs.~(\ref{forward_problem:eq3}) and (\ref{forward_problem:eq4}) was solved repeatedly until the iteration reached the minimum. 

For the forward problem, the computational domains were discretized with 
%30$\times$20 equilateral quad 
{\color{black}{square elements with 10~\mbox{mm} length on each side, which ensures that the obtained results have the same resolution along vertical and horizontal directions,}} and FEM has been applied for the {\color{black}spatial} discretization of the domain. 
External harmonic forcing was applied to the corresponding point on the FEM model with the amplitude of forcing being 1.0~N such that the resulting acceleration response can be regarded as the accelerance. 
For the implementation of the forward and inverse problems using FEM, COMSOL Multiphysics\textsuperscript{\textregistered} was used. 
The minimization problem was solved with sequential quadratic programming implemented as SNOPT solver in COMSOL. 
$\tilde{\bf H}(\omega_j)$ and ${\bf H}(\chi,\omega_j)$ in \eqref{eq4} or (\ref{eq5}) were evaluated at frequencies shown in \tabref{tab1}. To avoid the increase in the computational cost, only a limited number of frequencies were chosen. 
%These frequencies were chosen by the following criteria. The resonant frequencies were avoided, because they tend to move during the optimization and large deviations in the objective functions caused by the movement of the resonant frequencies sometimes caused the large change in the topology
% Conditions for the FE models.
Also note that when evaluating the FRFs by the FEA, the coefficients for the Rayleigh damping were set to $\alpha=0.421$~[rad/s] and $\beta=4.55\times10^{-6}$~[s/rad] for the test specimen A, and $\alpha=1.94$~[rad/s] and $\beta=7.53\times10^{-7}$~[s/rad] for the test specimen B, which were identified by preliminary experimental modal analyses on the specimens. 
%
% Conditions for the optimization 
For the objective function of the minimization problem defined in \eqref{eq2}, both $J(\chi)=J_1(\chi)$ and $J(\chi)=J_2(\chi)$ were examined. 
\begin{table}[tb]
\centering
\caption{Frequencies used for the minimization}\label{tab1}
\begin{tabular}{c|cccc}\hline
Case A1 & 60Hz & 140Hz & 350Hz & 450Hz\\
Case A2 & 145Hz& 325Hz& 450Hz& \\
Case B1 & 250Hz& 500Hz& 700Hz& \\
Case B2 & 260Hz& 804Hz& 900Hz& \\ \hline
\end{tabular}
\end{table}

\subsection{Results}
In this section, the results of the damage identification for the test specimens A and B are shown. In particular, the effects of Lasso regularization and the choice of the objective function on the results are discussed in detail. 

\subsubsection{Convergence history of the damage identification process}
\begin{figure}[bt]
\centering
%\subfigure[$J(\chi)=J_1(\chi)$, $\lambda=0.01$]{\includegraphics[width=8cm]{././Fig11a.pdf}}
%\subfigure[$J(\chi)=J_2(\chi)$, $\lambda=0.01$]{
\includegraphics[scale=1]{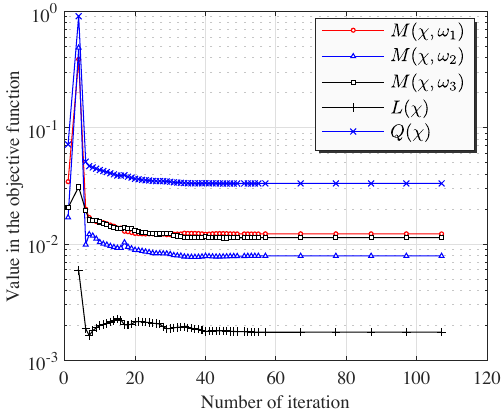}
%}
\caption{Convergence histories during the optimization process for Case B2 with $J(\chi)=J_2(\chi)$, $\lambda=0.1$}\label{inverse:fig5}
\end{figure}
First, to visualize the convergence of the proposed Algorithm \ref{alg1}, the convergence histories of the terms in the objective functions are examined. 
Representative convergence histories of the terms in the objective function corresponding to the damage identification for Case B2 with $J(\chi)=J_2(\chi)$ and $\lambda=0.1$ are shown in \figref{inverse:fig5}. 
As can be seen, $M(\chi,\omega_1)$, $M(\chi,\omega_2)$ and $M(\chi,\omega_3)$ greatly fluctuate for the first several iterations. After the large deviation, all terms gradually decrease as the optimization proceeds. It then terminates after the deviations in $Q(\chi)$ vanish. From these results, we can see that the optimization is successfully converged. Moreover, we can expect that the topology of the structure greatly changes for the first several iterations, and it does not change much afterwards. 
\begin{figure}[tb]
\centering
\includegraphics{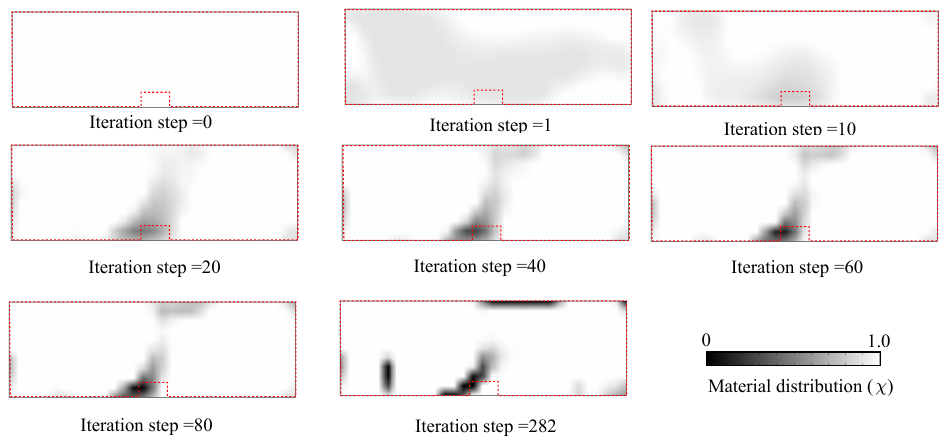}
\caption{Evolution of the material distribution for Case B2 with $J(\chi)=J_2(\chi)$ and $\lambda=0$. Domain boundary of the specimen is drawn with a dashed line.}\label{evol_fig1}
\end{figure}
\begin{figure}[bt]
\centering
\includegraphics{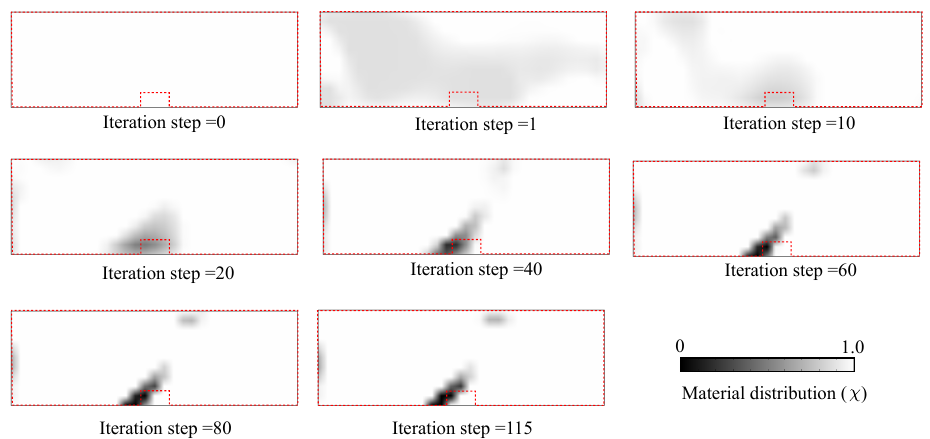}
\caption{Evolution of the material distribution for Case B2 with $J(\chi)=J_2(\chi)$ and $\lambda=0.1$ Domain boundary of the specimen is drawn with a dashed line.}\label{evol_fig2}
\end{figure}

Second, the evolutions of the topology are examined. Representative evolution histories of the topology are shown in \figref{evol_fig1} and \figref{evol_fig2} for Case B2 with $J(\chi)=J_2(\chi)$ and $\lambda=0$ and with $J(\chi)=J_2(\chi)$ and $\lambda=0.1$, respectively. For both figures, the domain boundaries of the specimen are drawn with dashed lines. 
Without the Lasso regularization, as can be seen in \figref{evol_fig1}, the algorithm first finds the rough estimate of the location of the damage by the 10th iteration step. Namely, the value of $\chi$ corresponding to the damaged region is reduced down to about 0.5 to 0.8. The value of $\chi$ then increases again for the areas where the damage does not exist, while that for the damaged region starts to decrease. 
The shape of the damage is then gradually refined as the optimization proceeds. 
At the same time, however, we can see that spurious damaged regions grow, which do not exist in the real specimen.  The spurious regions remain in the material distribution even when the termination criteria are satisfied at the 282nd iteration step. 
Note that the location of the rectangular notch is slightly off the identified region. 

With the Lasso regularization, it also finds the rough estimates of the location of the damage by 10th iteration step, as can be seen in \figref{evol_fig2}. The shape of the damage is then gradually refined. Notably, with the Lasso regularization, the generation of spurious regions is suppressed. The optimization terminates at the 115th iteration step, with the generation of slight spurious damaged regions.  
Again, the location of the rectangular notch is slightly off the identified area. However, the accuracy is greatly improved in comparison with the one shown in \figref{evol_fig1}, considering that the generation of spurious damaged regions is greatly suppressed. 

%The convergence histories of the terms for the representative results that employed $J(\chi)=J_1(\chi)$ and $J(\chi)=J_2(\chi)$ are shown in \figref{inverse:fig5}. 
%Note that the corresponding optimization results will be shown in \figref{inverse:fig2}(d) and \figref{inverse:fig4}(d) \figref{inverse:fig5}(a) and (b), respectively. 
%Even though the obtained topologies in \figref{inverse:fig2}(d) and \figref{inverse:fig4}(d) are qualitatively similar to each other, the convergence histories shown in \figref{inverse:fig5}(a) show slightly slower convergence rate than those in \figref{inverse:fig5}(b). This may also show the advantage of using the objective function $J_2(\chi)$. 

\clearpage
\subsubsection{The effects of the choice of objective functions and Lasso regularization on the identification results}
\begin{figure}[tb]
\centering
\includegraphics[scale=1]{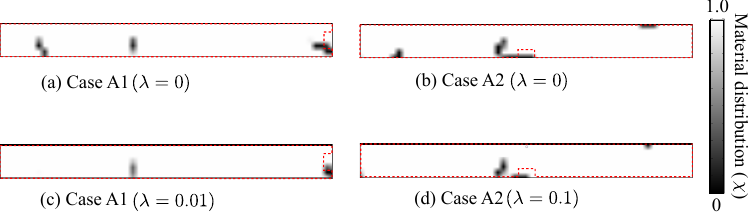}
\caption{Results of the damage identification for the test specimen A with $J(\chi)=J_1(\chi)$. Domain boundary of the specimen is drawn with a dashed line.}\label{inverse:fig1}
\end{figure}

First, the damage identification results for the test specimen A with $J(\chi)=J_1(\chi)$ are shown in \figref{inverse:fig1}. Namely, the damage identification was conducted by minimizing the pointwise error between the measured and the computed FRFs. 
Figures \ref{inverse:fig1}(a) and (b) show the results without Lasso, or $\lambda=0$, whereas figures \ref{inverse:fig1}(c) and (d) show those with Lasso. 
As can be seen in \figref{inverse:fig1}(a), reduction in the material distribution near the short edge can be observed, which indicates that there is a void region, or the damage in the real structure. 
However, the location of the identified damaged region is slightly off the place where the notch exists in the real structure. Furthermore, we can see that there are spurious damaged areas near the fixed edge and in the middle of the plate. 
The same trend can be seen in \figref{inverse:fig1}(b), where the reduction in the material distribution can be observed in the damaged regions, and some spurious damaged regions appear. 
These results indicate that the solution to the minimization problem is prone to spurious damaged regions. 
With the Lasso regularization, some of the spurious damaged regions that appear in \figref{inverse:fig1}(a) and (b) disappear. This shows the effectiveness of the regularization. However, we can also see that the area of the real damaged regions is also reduced due to the regularization. 

\begin{figure}[tb]
\centering
\includegraphics[scale=1]{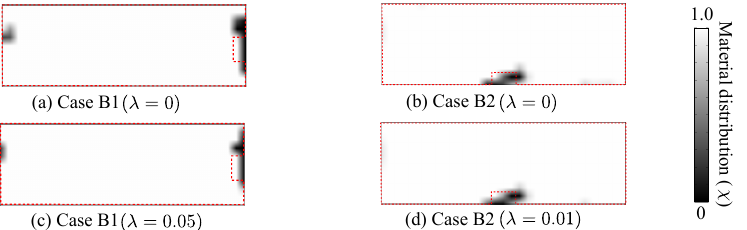}
\caption{Results of the damage identification for the test specimen B with $J(\chi)=J_1(\chi)$. Domain boundary of the specimen is drawn with a dashed line.}\label{inverse:fig2}
\end{figure}
Second, the damage identification results for the test specimen B with $J(\chi)=J_1(\chi)$ are shown in \figref{inverse:fig2}. As can be seen in \figref{inverse:fig2}(a) and (b), the damaged regions were successfully identified by the proposed method. However, the shape of the notch was not correctly identified. In comparison with the results for the test specimen A, there are less spurious damaged regions for this specimen, even without the regularization. 
As can be seen in \figref{inverse:fig2}(c) and (d), the identified damaged regions without the regularization become small with the Lasso regularization. This was also the case with the results for the test specimen A. 
\begin{figure}[tb]
\centering
\includegraphics[scale=1]{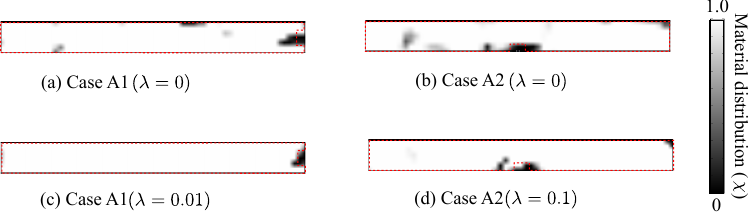}
\caption{Results of the damage identification for the test specimen A with $J(\chi)=J_2(\chi)$. Domain boundary of the specimen is drawn with a dashed line.}\label{inverse:fig3}
\end{figure}

Third, the damage identification results for the test specimen A with $J(\chi)=J_2(\chi)$ are shown in \figref{inverse:fig3}. That is, the correlation between the measured and the computed FRFs is maximized, or its negative is minimized. 
As can be seen in \figref{inverse:fig3}(a) and (b), without the regularization, there are more spurious damaged areas than those observed in the results with $J(\chi)=J_1(\chi)$ shown in \figref{inverse:fig1}(a) and (b). This difference is caused by the nature of $J_2(\chi)$ where the correlation of the FRFs is maximized, as opposed to that of $J_1(\chi)$ where the point-wise errors are minimized. However, the damaged areas are correctly identified for both Cases A1 and A2. 
With the regularization, the number of the spurious damaged areas is greatly reduced, as can be seen in \figref{inverse:fig3}(c) and (d). Even though the areas of the damaged regions are reduced by the regularization, we can clearly see the damaged regions from the material distribution. 

\begin{figure}[tb]
\centering
\includegraphics[scale=1]{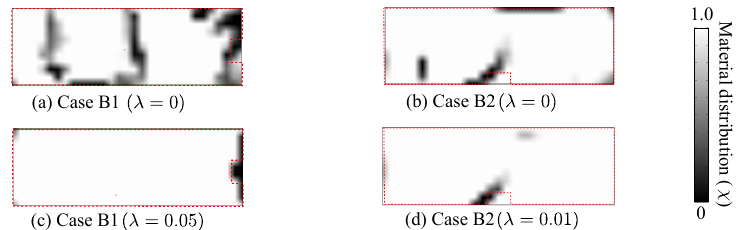}
\caption{Results of the damage identification for the test specimen B with $J(\chi)=J_2(\chi)$. Domain boundary of the specimen is drawn with a dashed line.}\label{inverse:fig4}
\end{figure}
Finally, the damage identification results for the test specimen B with $J(\chi)=J_2(\chi)$ are shown in \figref{inverse:fig4}. As can be seen in \figref{inverse:fig4}(a) and (b), there are many and large spurious damaged regions in the results. The same trend observed in the results for the test specimen A with $J(\chi)=J_2(\chi)$ in \figref{inverse:fig3}(a) and (b) can also be observed for this results. This time the spurious damaged regions are so large that the real damaged regions cannot really be identified from the results. 
With the regularization, however, the number of the spurious damaged regions is greatly reduced, as can be seen in \figref{inverse:fig4}(c) and (d). The proposed method still cannot determine the accurate area of the damage well, but we can see that the location and shape of the damage can be estimated from those results. 

Overall, the results with $J(\chi)=J_1(\chi)$ are better without the regularization, but due to the strong nature of the objective function, the optimization algorithm produces spurious damaged regions even with the regularization. 
On the other hand, the results with $J(\chi)=J_2(\chi)$ show that even though many spurious damaged regions are generated without the regularization, they can be relatively easily removed by the regularization. This may be attributed to the weak nature of the objective function. 

{\color{black}{As a final remark, it is noted that to determine the proper value of $\lambda$, parametric studies with respect to $\lambda$ need to be conducted with known damage topologies and the value of $\lambda$ needs to be optimized. This can also be done effectively with numerically-obtained reference data with controlled noise levels, and may be automated by considering another optimization problem with respect to $\lambda$. This way, the proposed method can be extended to other structures and damage topologies.}}

\section{Conclusions}\label{sec:conclusions}
In this paper, a novel damage identification method has been proposed, which identifies the damaged regions in the structure as regions where material does not exist, by minimizing the differences between experimentally-obtained FRFs and those computed by FEAs. The minimization problem was formulated as a topology optimization with the SIMP method. The key idea is to add the regularization term in the objective function, which penalizes the $L_1$ norm of the material distribution. This greatly reduces the generation of spurious damaged regions. 

The proposed method was applied to the experimentally-obtained FRFs of representative structures with a notch that imitates material loss or the damage. The proposed method has been applied to four representative cases, and it was shown that it can successfully identify the location and rough estimates of the shape of the notch, especially when the regularization is applied. 

We provide the FRF dataset of the damaged test specimens used in this study via the Internet.
The dataset can be used as a benchmark problem for vibration-based damage detection methodologies.

{\color{black}{
The main contributions of this paper are summarized as follows. 
\begin{itemize}
\item  Experimental validation of the damage identification based on vibration data and topology optimization is given, which, to the knowledge of the authors, has never been done previously. 
\item It is shown that the proposed method is capable of identifying the damage with a reasonable accuracy, even with noisy experimental data. Most of the previous works dealt with data under idealized situations without measurement noises. 
\end{itemize}
With these contributions, the authors believe that the applicability of the proposed method to practical engineering problems is greatly increased. 
}}
\section{Acknowledgements}
This project was supported in part by Japan Society for the Promotion of Science (JSPS), Grant-in-Aid for Scientific Research(C), grant number 20K11855. The support of JSPS is gratefully acknowledged.
%\bibliographystyle{elsarticle-num}
%\bibliography{../../../References/references}

%
%\clearpage
%\listoftables
%\listoffigures
\end{document}